\documentclass[a4paper]{article}

\usepackage{INTERSPEECH2021}
\usepackage{graphicx}
\usepackage{float}
\usepackage{booktabs}
\usepackage{xcolor}
\usepackage{color}
\usepackage{hyperref}

\usepackage{subcaption}

\usepackage{caption}
\usepackage{enumitem}

\title{FoundationTTS: Text-to-Speech for ASR Customization \\with Generative Language Model}
\name{Ruiqing Xue$^{*}$, Yanqing Liu$^{*}$, Lei He, Xu Tan, Linquan Liu, Edward Lin, Sheng Zhao}
\address{Microsoft}
\email{}

\begin{document}

\maketitle
\def\thefootnote{*}\footnotetext[1]{These authors contributed equally to this work. Corresponding author: Ruiqing Xue, ruiqingxue@microsoft.com}\def\thefootnote{\arabic{footnote}}
%\def\thefootnote{*}\footnotetext[1]{Corresponding author: Yanqing Liu, yanqliu@microsoft.com}\def\thefootnote{\arabic{footnote}}
%\footnote{normal footnote}

\begin{abstract}
Neural text-to-speech (TTS) generally consists of cascaded architecture with separately optimized acoustic model and vocoder, or end-to-end architecture with continuous mel-spectrograms or self-extracted speech frames as the intermediate representations to bridge acoustic model and vocoder, which suffers from two limitations: 1) the continuous acoustic frames are hard to predict with phoneme only, and acoustic information like duration or pitch is also needed to solve the one-to-many problem, which is not easy to scale on large scale and noise datasets; 2) to achieve diverse speech output based on continuous speech features, complex VAE or flow-based models are usually required. In this paper, we propose FoundationTTS, a new speech synthesis system with a neural audio codec for discrete speech token extraction and waveform reconstruction and a large language model for discrete token generation from linguistic (phoneme) tokens. Specifically, 1) we propose a hierarchical codec network based on vector-quantized auto-encoders with adversarial training (VQ-GAN), which first extracts continuous frame-level speech representations with fine-grained codec, and extracts a discrete token from each continuous speech frame with coarse-grained codec; 2) we jointly optimize speech token, linguistic tokens, speaker token together with a large language model and predict the discrete speech tokens autoregressively. Experiments show that FoundationTTS achieves a MOS gain of +0.14 compared to the baseline system. In ASR customization tasks, our method achieves 7.09\% and 10.35\% WERR respectively over two strong customized ASR baselines.

\end{abstract}
%\end{}

\section{Introduction}
\label{sec_intro}

Text-to-speech (TTS) can help reduce word error rate (WER) in the customization of automatic speech recognition (ASR) especially for low resource or out-of-vocabulary (OOV) scenarios by generating synthetic speech, and using the text and speech pairs to update the ASR models or using a spelling correction model with hypothesis decoded from TTS data~\cite{li2020developing,sim2019personalization,wang2021light,wang2022towards,wang2023improving,papamakarios2021normalizing,ueno2019multi,tjandra2018machine,mimura2018leveraging}. However, compared to real speech in ASR training set, the adaptation performance with synthesized speech is inferior due to a lack of complex recording conditions, diverse and realistic speaker variance, and dynamic prosodic speech for different linguistic contexts, which reveals the drawback of neural TTS that is trained with relatively small and clean dataset compared to ASR applications that usually have 1000x more data. Thus, the closer the synthesized speech can simulate compared to real speech in the ASR training set, the better the customization performance. TTS models with a large-scale data set and diverse speech output are the key factors for better ASR customization with synthesized speech.

Neural text-to-speech synthesis~\cite{wang2017tacotron,shen2018natural,li2019neural,ren2019fastspeech, li2020robutrans,ren2021fastspeech,liu2021delightfultts,donahue2020end, liu2022delightfultts,kim2021conditional,weiss2021wave,tan2021survey,tan2022naturalspeech} has shown close to human quality in terms of naturalness and audio fidelity in the past years, with cascaded or end-to-end Transformer or LSTM architectures. Cascaded models separately optimize acoustic model and vocoder with autoregressive~\cite{li2019neural} or non-autoregressive~\cite{liu2021delightfultts} Transformer layers, while end-to-end (E2E) TTS simplifies the model building by jointly training acoustic model and vocoder to improve the speech quality with traditional mel-spectrograms ~\cite{ren2021fastspeech} or automatically learned speech representations~\cite{liu2022delightfultts}. The predicted frame-level features are transformed to raw waveform by the neural vocoders with adversarial training~\cite{kumar2019melgan,kong2020hifi,kim2021fre, jang2021univnet}. DelightfulTTS 2~\cite{liu2022delightfultts} proposes a new codec network based on vector-quantized auto-encoders with adversarial training to automatically learn frame-level speech representations. Instead of using mel-spectrograms or other pre-designed features, it uses the codec encoder to extract speech representations, quantize them with residual vector quantizers, and then uses the codec decoder to reconstruct waveform with adversarial training. Though the above works achieve high quality on a small and clean dataset, it is hard to scale with larger or noisy datasets with diversity control due to complex and continuous acoustic feature dependency between the acoustic model and vocoder.

TTS with continuous speech feature is hard to solve the one-to-many problem as output randomness is hard to implement on frame-level speech features to achieve observable diversity on prosody and speaker, and features like mel-spectrograms, linear-spectrograms or self-extracted speech features are highly correlated along time and frequency axes which leads to overfitting or over smoothing effects for the acoustic model to predict. Human speech naturally has a random sampling effect like white noise in source-filter vocoders~\cite{hu2013experimental,zen2007hmm,kawahara1999restructuring, drugman2011deterministic,tokuda1994mel} that linguistic or acoustic condition is not enough to simulate. Since diverse control is easier to achieve than language or image models do, some works leverage discrete speech tokens to explore different architectures to replace continuous feature-based models. ~\cite{hayashi2020discretalk} proposes an E2E TTS framework based on VQ-VAE ~\cite{razavi2019generating}
and neural machine translation (NMT), where the VQ-VAE model learns a mapping from the speech waveform into a sequence of discrete tokens, and the Transformer-based NMT model is trained to estimate the discrete token sequence from a given input, due to the use of discrete speech tokens, it can do sampling at each step as in NMT and ASR and thus avoid the over smoothing problem. VQTTS~\cite{du2022vqtts} consists of an acoustic model (AM) called txt2vec and a vocoder called vec2wav, which uses a self-supervised vector-quantized (VQ) acoustic feature rather than mel-spectrograms, in which txt2vec is a classification model with cross-entropy loss rather than a regression model while vec2wav uses an additional feature encoder for smoothing the discontinuous quantized feature. Instead of predicting the complicated mel-spectrograms, txt2vec only needs to consider the correlation along the time axis, which reduces the gap between ground truth speech frames and the predicted ones. Although diversity can be achieved by sampling in inference time, these discrete models typically use a very small amount and relatively clean dataset like popular JUST or LJ Speech ~\cite{sonobe2017jsut,ljspeech17}, but large-scale and nosier datasets are not yet systemically explored especially in diverse speech output for ASR customization.

Large language model (LLM) based acoustic model is making remarkable progress on large-scale speech dataset training. Speech recording typically contains acoustic or semantic abstractions like fine-grained phonetic-level or word-level prosody, syntax, grammar attributes, or coarse-grained attributes like speaker identification, style class, accent type, etc. Neural codec models have achieved close to recording quality by leveraging methods such as autoregressive waveform modeling~\cite{oord2016wavenet,valin2019lpcnet,kalchbrenner2018efficient} and adversarial training~\cite{zeghidour2021soundstream,liu2021delightfultts,tagliasacchi2020seanet, defossez2022high}. Language models have also demonstrated their ability to model high-level, long-term sequences for different content types, as shown by the recent advances in text~\cite{brown2020language,radford2019language,ouyang2022training,wei2021finetuned, chowdhery2022palm} and image~\cite{yu2021vector,ramesh2022hierarchical,ramesh2021zero,he2022masked,bao2021beit,lee2022autoregressive}. Audio generation~\cite{borsos2022audiolm,kreuk2022audiogen,lakhotia2021generative, kreuk2022audio,maiti2022speechlmscore,maimon2022speaking} has made progress towards a natural audio synthesis that remains natural and consistent over time. ~\cite{borsos2022audiolm} shows that a Transformer~\cite{vaswani2017attention} decoder trained on discretized speech units can generate coherent speech without relying on textual phonemes, by a continuing generation with a speech prompt or music prompt sequence with dynamic acoustic diversity, linguistic consistency, and high audio fidelity. However, the prompt is speech unit only and the speaker or text prompt is not yet explored.

In this work, we propose FoundationTTS, a framework that enables high-quality and diverse text-to-speech generation with long-term linguistic and acoustic dependency modeling on large-scale speech datasets. As demonstrated by our experiments, the speech output of FoundationTTS is more diverse than traditional models due to discrete sampling mechanisms in language models. We achieve this objective by combining recent advances in the adversarial neural audio codec, end-to-end neural text-to-speech~\cite{liu2022delightfultts}, and large language modeling~\cite{brown2020language}. Specifically, starting from a raw audio waveform, we first reconstruct the waveform with a fine-grained audio codec, where the codec encoder extracts the frame-level speech representations from the waveform with a set of residual quantizers, which capture the acoustic details of the audio waveform and allow for high-quality reconstruction by the codec decoder. Then we construct another coarse speech codec to reconstruct the fine-grained speech representations with few quantizers to get the acoustic/speech tokens. At last, we trained a decoder-only language model, which autoregressively predicts the acoustic tokens from the coarse-grained speech codec which captures both local dependencies (e.g., phonetics in speech) and global long-term structure (e.g., semantic content in speech). Modeling both linguistic and acoustic tokens lead to both high audio quality and long-term consistency. In summary, we make the following contributions: 
\begin{itemize}[leftmargin=*]
\item We propose FoundationTTS, a framework for text-to-speech generation, which combines discrete speech tokens in a hierarchical fashion with a large language model to achieve long-term consistency and diverse output on large-scale and complex speech datasets. 
\item We propose a hierarchical codec network with the fine-grained codec reconstructing waveform with frame-level speech representations and coarse-grained codec reconstructing the speech representations with fewer quantizers to convert continuous representations to discrete speech tokens for language modeling. 
\item We demonstrate the ability of FoundationTTS to generate linguistically consistent, prosody-diverse, and high-quality speech for ASR domain customization and general neural text-to-speech.
\end{itemize}

\begin{figure*}[h]
  \centering
  \includegraphics[width=0.8\textwidth,trim=0.5cm 0.5cm 0.5cm 0.5cm,clip=true]{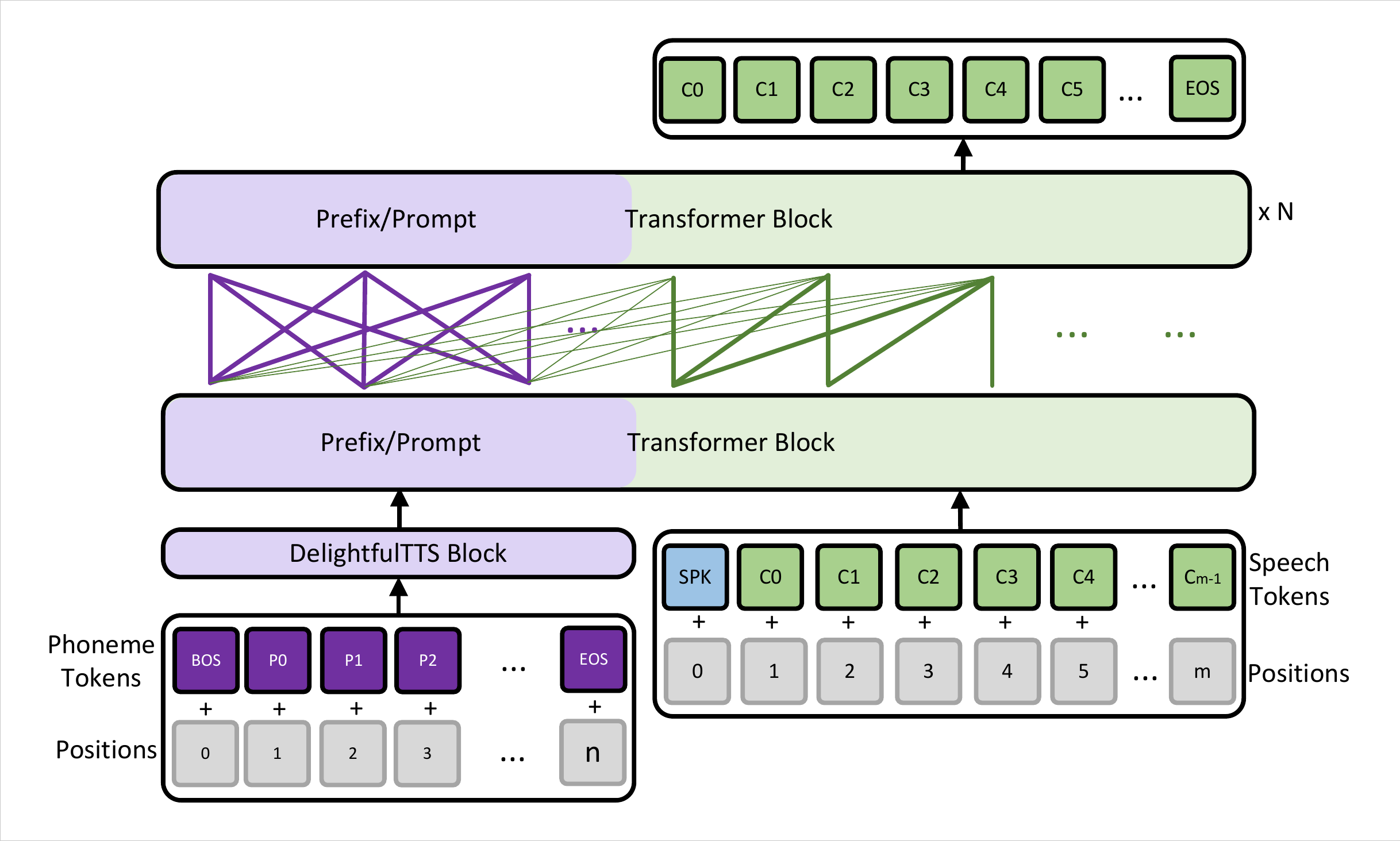}
  \caption{Prefix language model of FoundationTTS. DelightfulTTS blocks act as the phoneme encoder with phoneme tokens and position embedding as input. A speaker token is also added to the prefix noted as SPK. A set of transformer decoders perform autoregressive decoding of speech tokens with full phoneme tokens and speaker tokens.}
  \label{fig:lm}
\end{figure*}

\section{FoundationTTS}
\label{sec_overview}
FoundationTTS is a two-stage TTS model, in the first stage, we introduce a hierarchical audio code that can encode and reconstruct waveform, more specifically, it consists of two components: 1) fine-grained audio codec for continuous speech frame extraction. 2) coarse-grained audio codec for discrete speech token extraction. In the second stage, a prefix language model is trained over the discrete speech token sequence generated by the audio codec in the first stage. We use the prefix language model instead of the encoder-decoder paradigm because it can simplify the model structure and has gained great success in GPT-like models.

\subsection{Speech Tokenizer with Hierarchical Audio Codec}
\label{sec_codec}

\begin{figure*}[h]
  \centering
  \includegraphics[width=0.93\textwidth,trim=0.5cm 0.5cm 0.5cm 0.5cm,clip=true]{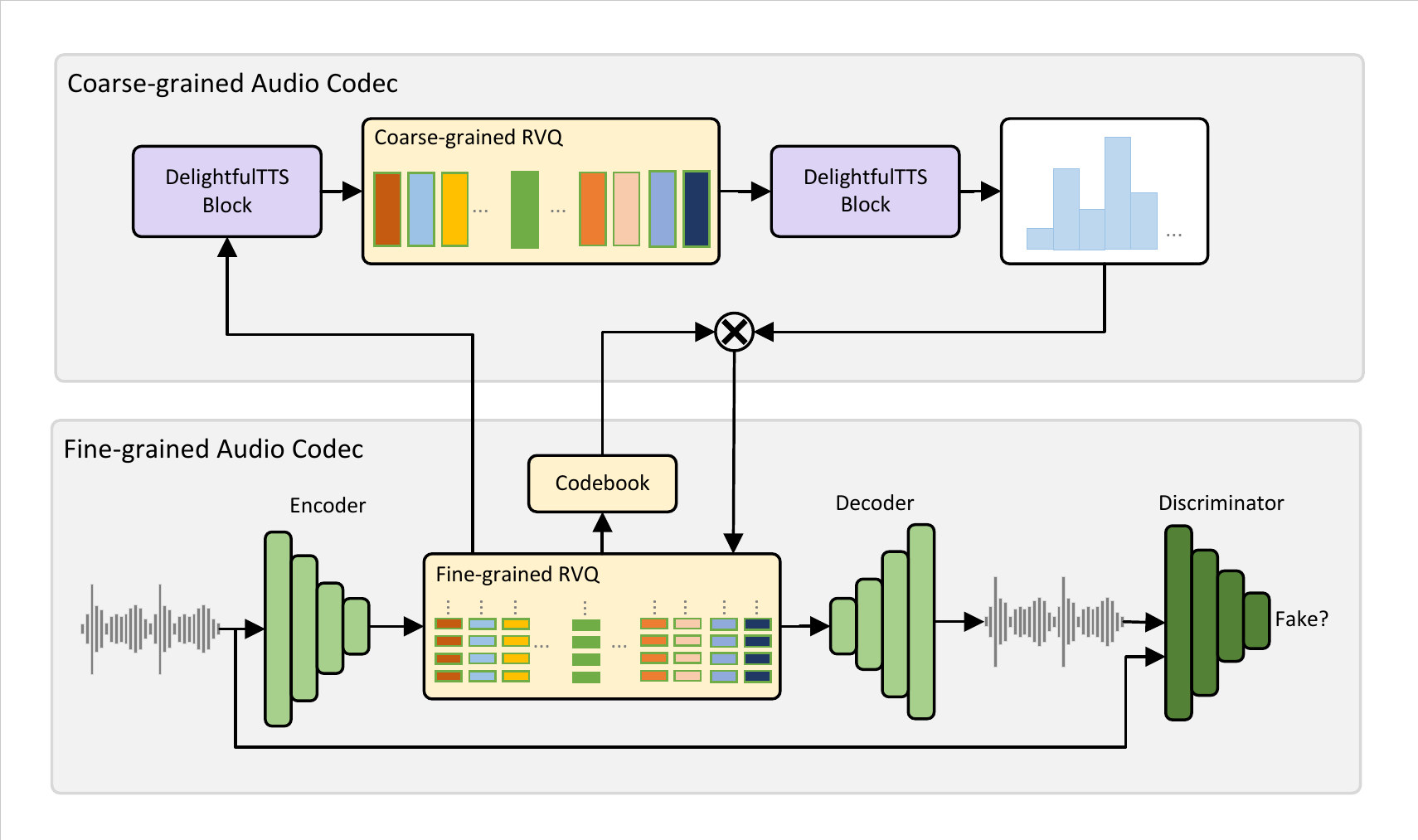}
  \caption{The structure of the hierarchical codec network consists of a fine-grained codec and a coarse-grained codec. The fine-grained codec is composed of a convolution down-sampling encoder that extracts frame-level speech representations from audio waveform, an RVQ block that discretizes the frame representation, and a convolution up-sampling decoder. The structure of coarse-grained codec is like fine-grained codec except that it uses DelightfulTTS block instead of convolution network and the number of RVQ layers is 1.}
  \label{fig:hcodec}
\end{figure*}

The neural audio codec uses more bitrate to reconstruct high-quality waveform comparable to real speech with discrete tokenizers at the frame level, which will increase the discrete speech token number at each frame with multiple residual quantizers but directly using smaller quantizer is hard to reconstruct waveform because real audio samples are complex with noises, besides language model performs poor on too long-range context modeling and the tokenizer is based on character or word piece which is much shorter than speech frames. The fine-grained codec frame and coarse-grained LM model unit thus need a balance for audio codec reconstruction quality and LM performance. To solve this contradiction, as in Figure~\ref{fig:hcodec}, we designed a hierarchical audio codec as a speech tokenizer: first, the fine-grained audio codec encodes waveform to continuous features with a convolutional encoder and reconstructs waveform with a mirror decoder from the bottleneck layer via residual quantizers; then the coarse-grained audio codec further compresses the continuous representation generated by the fine-grained codec using fewer quantizers for shorter speech token sequence.

\subsubsection{Fine-Grained Codec with VQGAN}
%\noindent\textbf{
Fine-grained codec is like the frame-level feature extraction network in DelightfulTTS 2~\cite{liu2022delightfultts}, which consists of a waveform down-sampling encoder, and another mirrored up-sampling decoder, and a residual vector quantizer acting as a feature bottleneck. The codec encoder receives 16kHZ waveform as input, and down-sampling it to frame-level representation by convolutional layers, resulting in a 600X compression ratio with down-sampling layers. The frame-level speech representation is then quantized by a residual vector quantizer (RVQ) with multiple vector quantizers; thus, we can compress the 16kHz continuous waveform samples into 26.7 Hz discrete speech tokens with 3.4 kbps, with a frame length of 37.5ms, in this way we can tokenize each speech frame into multiple discrete tokens, but it’s slow to predict multiple discrete tokens in one step like ~\cite{borsos2022audiolm}, which flattens multiple speech tokens at each step and predicts autoregressively from coarse quantizer layers.

\subsubsection{Coarse-Grained Codec for LM Token Extraction}
A coarse-grained codec is designed to reconstruct the frame level feature from the fine-grained codec but with fewer quantizers, more quantizers will better reconstruct waveform in a fine-grained codec but it will also increase the speech token number for one speech frame and make it hard for language modeling with discrete tokens, the coarse-grained codec makes a balance on the token number and waveform reconstruction quality based on the pre-trained fine-grained codec. The frame-level acoustic feature from the fine-grained codec acts as the reconstruction target for the coarse-grained codec. The fine-grained frame-level speech acoustic features are first fed into an improved conformer encoder~\cite{liu2021delightfultts}, and the output is then quantized by a single-layer vector quantizer. Finally, the fine-grained audio codec acoustic features are reconstructed by another conformer decoder. The purpose of reconstructing the fine-grained speech frames instead of the waveform is to simplify the predicting target and thus we can use fewer quantizers to approximate the speech by the fine-grained codec. To further improve reconstruction ability, we don’t reconstruct features directly, instead, the decoder predicts the weight distribution of fine-grained RVQ, which we found will perform better tokenization in experiments and speed up training convergence.

We used the same discriminators described in Delightfultts 2 ~\cite{liu2022delightfultts}, which combine multi-scale and multi-period discriminators for both fine-grained audio codec and coarse-grained audio codec.

\subsection{TTS Language Modeling}
\label{sec_llm}
The acoustic model in traditional neural TTS is designed to predict continuous speech features, for example, mel-spectrograms used by TransformerTTS, FastSpeech, and DelightfulTTS~\cite{li2019neural, ren2019fastspeech,liu2021delightfultts} or self-learned acoustic features used by NaturalSpeech and DelightfulTTS 2~\cite{tan2022naturalspeech,liu2022delightfultts} from linear spectrum or waveform directly. Although these models can synthesize high-quality and natural speech, two drawbacks are still obvious: 1) hard to leverage complex real-world speech or large-scale speech datasets for acoustic modeling, such as noisier data in ASR corpus or audiobook instead of relatively clean recorded speech. 2) it is not straightforward to synthesize diverse speech output from continuous frame-level features to reduce the one-to-many mapping issue in TTS. 
Generally, optional approaches are needed to alleviate the diverse issue. The first class is to provide additional acoustic or linguistic conditions for the acoustic model except for the phoneme sequence input, such as pitch or phoneme embeddings~\cite{lancucki2021fastpitch,wang2018style, hsu2018hierarchical}, prosodies in different granularities have also been added to TTS in ~\cite{sun2020fully,chien2021hierarchical}. 
The effectiveness of VQVAE is also explored in~\cite{wang2019vector}
. Another approach to address the problem is to exploit better training criteria except for the common L1 or L2 loss criterion for AM. However, the real acoustic distribution is much more complicated. Hence, some research uses normalizing flow ~\cite{papamakarios2021normalizing,kim2020glow} for this problem. The normalizing flow transforms the data distribution into a known simple distribution and is optimized via maximum log-likelihood. However, both VAE and flow models should be carefully designed to ensure invertibility and stability, which restricts the scalability of such models.

As in Figure~\ref{fig:lm}, to overcome these problems, we use discrete speech tokens as the modeling unit and designed a prefix language model to predict these discrete tokens. Specifically, the speech-language model consists of a DelightfulTTS encoder block and a decoder-only Transformer block. The DelightfulTTS encoder extracts linguistic representations from phoneme or text sequence, and the decoder-only transformer will autoregressively generate discrete acoustic tokens with linguistic representations as a prefix. The linguistic prefix and discrete speech tokens are fed into the same transformer decoder without a cross-attention layer. Different position embeddings and tokens are used to distinguish them from each other. We use bi-directional self-attention over linguistic prefixes and use monotonous self-attention on discrete speech tokens. Speaker embeddings are also added as speaker prefixes in the Transformer decoder. It’s observed that larger LM is more stable than smaller models in terms of naturalness or bad case ratio like skipping phone or repeat phone issues.

\begin{figure*}[h]
  \centering
  \includegraphics[width=0.93\textwidth,trim=0.5cm 0.5cm 0.5cm 0.5cm,clip=true]{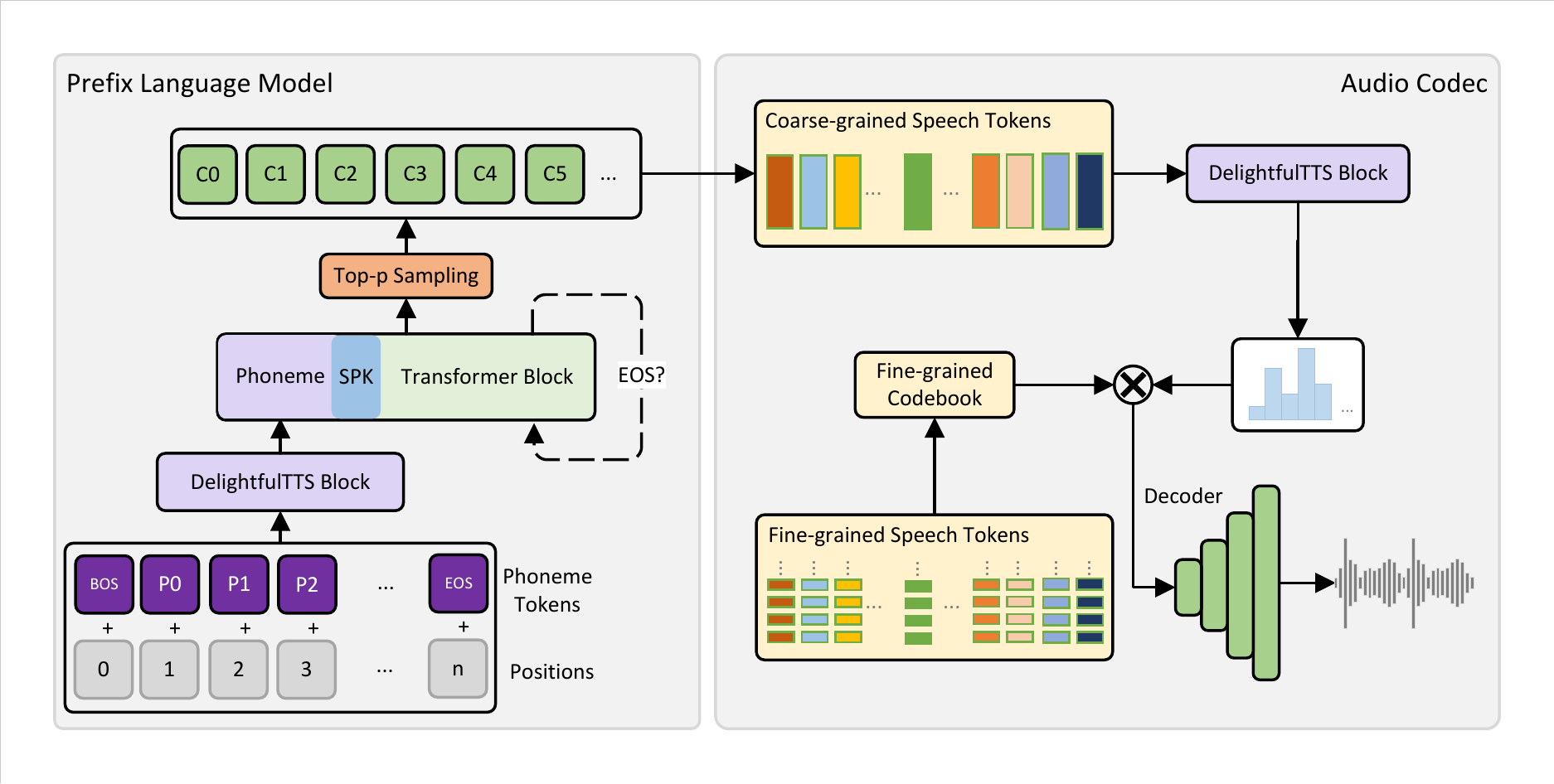}
  \caption{During inference, phoneme tokens are first fed into a DelightfulTTS encoder to extract the linguistic representation, and a decoder-only Transformer autoregressively generates speech tokens conditioned on these linguistic features and speaker token, we apply top-p sampling to generate diverse speech. Finally, the audio waveform will be reconstructed by hierarchical codec decoders based on the predicted speech tokens.}
  \label{fig:inference}
\end{figure*}

\subsection{Training Objectives}
\label{sec_obj}

\textbf{Codec Loss} Following DeflightuflTTS 2, we train the hierarchical audio codec with adversarial objectives. 

\begin{equation}
L_{\rm FCODEC} = L_{\rm adv}+L_{\rm rvq}+L_{\rm fm}+L_{\rm mrs}+L_{\rm msd}+L_{\rm mpd}
\end{equation}

Fine-grained codec loss contains a multi-period discriminator loss $L_{\rm mpd}$, multi-scale discriminator loss $L_{\rm msd}$, together with an adversarial loss $L_{\rm adv}$, a multi-resolution spectrogram loss $L_{\rm mrs}$, a feature match loss $L_{\rm fm}$ and residual vector quantization loss $L_{\rm rvq}$ with random segmented waveform.

\begin{equation}
L_{\rm CCODEC} = L_{\rm adv}+L_{\rm vq}+L_{\rm fm}+L_{\rm mrs}+L_{\rm msd}+L_{\rm mpd}
\end{equation}

The coarse-grained codec uses the reconstructed fine-grained frames, goes through the frozen fine-grained speech codec and generates a reconstructed waveform with the same multi-period discriminator loss $L_{\rm mpd}$, multi-scale discriminator loss $L_{\rm msd}$, together with an adversarial loss $L_{\rm adv}$, a multi-resolution spectrogram loss $L_{\rm mrs}$, a feature match loss $L_{\rm fm}$ and vector quantization loss $L_{\rm vq}$.

\noindent\\\textbf{Prefix Language Model Loss} given an audio waveform $x = \left \{ x_{1},...,x_{t_{x}} \right \} $, the hierarchical codec compresses it into discrete speech tokens as LM target, $c = \left \{ c_{1},...,c_{t_{c}} \right \} $, and a DelightfulTTS block is used to map phoneme sequences into linguistic representations $w = \left \{ w_{1},...,w_{t_{w}} \right \} $, then the speech synthesis task can be viewed as a language model problem that trained with cross-entropy loss.

\begin{equation}
L_{LM}=-\mathbb{E} \sum_{t=1}^{T_{c}} \log_{}{p_{\theta }\left ( c_{t} |c_{1},...,c_{t-1},w_{1},...,w_{t_{w}}\right ) } 
\end{equation}

Where $T_{c}$ is the frame number and $t_{w}$ is the length of the phoneme sequence, ${\theta }$ is the trainable parameters of the language model.

\subsection{Inference}
During inference, as in Figure~\ref{fig:inference}, the input phoneme sequence is first fed into a DelightfulTTS block in the prefix LM to extract the linguistic representation, and the decoder-only transformer autoregressively generates speech tokens conditioned on these linguistic features and speaker embedding. Another DelightfulTTS block in a coarse-grained codec decoder converts the discrete speech tokens into a frame-level weight distribution of RVQ tokens in the fine-grained codebook, and then the corresponding fine-grained speech tokens are generated based on the distribution. Finally, the fine-grained codec decoder reconstructs the waveform based on these fine-grained speech tokens. We apply top-p sampling to generate diverse speech.

\section{Experiments}
\label{sec_exp}
We explain the experiment dataset, training, and evaluation setting for the proposed FoundationTTS with TTS and ASR metrics in this section. 

\subsection{Experimental Setup}
\noindent\textbf{Datasets} to evaluate the performance on ASR customization, we conduct experiments on both internal Chinese shanghai dialect datasets and English datasets. The internal Chinese shanghai dialect dataset contains 600 hours of the domain-specific corpus, the transcriptions are converted into phonemes by the rule-based grapheme-to-phoneme module, since the transcription contains polyphonic characters and the rule-based G2P module cannot accurately disambiguate, the polyphonic characters are reserved as extra tokens. The English datasets contain 900 hours corpus, which consists of LibriTTS, VCTK, and internal TTS data ~\cite{zen2019libritts,veaux2016superseded}, the text goes through the standard TTS front-end data processing process including rule-based text normalization and grapheme-to-phoneme conversion. The sample rate of the audio waveform is 16kHz.

\noindent\\\textbf{Training Setting} FoundationTTS consists of three components that are trained one by one, firstly, the fine-grained audio codec is trained with audio-only data. Secondly, the coarse-grained audio codec is trained with the pre-trained fine-grained audio codec. Finally, the prefix language model is trained with the speech token extracted with the coarse-grained audio codec.  8x 32GB V100 GPUs are used for both hierarchical codec and prefix language model training. During hierarchical codec training, we randomly segment the waveform into chunks of 24000 sample points to speed up model convergence. Adam optimizer is used for hierarchical codec and prefix language model training, with Radam and Lookahead optimization~\cite{liu2019variance,zhang2019lookahead}; for prefix language model training, we apply a learning rate of 0.001 with exponentially decaying to 0.00005 starting from 4,000 iterations. The fine-grained codec contains four convolutional layers in the encoder and decoder, and 16 residual quantizers are applied in between. The coarse-grained contains three improved conformer blocks in the encoder and decoder and one quantizer is applied after the codec encoder. The prefix LM contains 16 Transformer layers, and six improved conformer blocks as phoneme encoders, with a total parameter number of about 600M.

\noindent\\\textbf{Evaluation Setting} We evaluate the adaptation performance FoundationTTS on two independent ASR systems, Hybrid, and RNN-T~\cite{graves2012sequence,rao2017exploring} on both Chinese shanghai dialect and English test set with word error rate (WER) metrics. We also evaluate the synthesized audio quality of FoundationTTS and reconstruct the quality of the hierarchical codec with both objective and subjective metrics, subjective metrics include the mean opinion score (MOS) of FoundationTTS and comparative mean option score (CMOS) of codec, ViSQOL ~\cite{hines2015visqol} score of coarse-grained audio codec and fine-grained audio codec is reported as the objective metrics. The TTS and codec evaluation set used by subjective and objective metrics evaluation are randomly preserved from the training set, and the ASR adaptation evaluation set does not overlap with the training set.

\subsection{Results}
\label{sec_test_asr}
\subsubsection{ASR Adaptation Performance}

\textbf{Hybrid ASR adaptation} Table~\ref{tab:ShanghaiDialect} shows the WER reduction on the hybrid ASR system with FoundationTTS domain-specific adaptation synthesized data, the baseline hybrid system is a pre-trained Chinese shanghai dialect model, note that the baseline model is already adapted with domain-specific data, even so, our model still achieves a 7.09\% WER reduction, the evaluation set contains 1000 sentences.

\begin{table}[h]
  \centering
  %\resizebox{\linewidth}{!}{%
  \begin{tabular}{llll}
    \toprule
     & Female & Male & ALL  \\
    \midrule
    Hybrid Baseline  & 9.96\% & 10.63\%  & 10.29\%   \\
    \midrule
    +FoundationTTS adaptation & 8.88\% & 10.26\% & 9.56\%  \\
    \midrule
    WERR  & 10.84\% & 3.48\% & 7.09\%  \\
    \bottomrule
  \end{tabular}
  \caption{ WERR of ASR adaptation on Shanghai Dialect test set.}
  \label{tab:ShanghaiDialect}
\end{table}

\noindent \textbf{RNN-T adaptation} We conduct another FoundationTTS adaptation experiment on the RNN-T ASR model, for comparison, we also list the WER of FastSpeech adaptation. FoundationTTS adaptation and FastSpeech adaptation are trained with the same domain-specific dataset. Table~\ref{tab:RNN-T} shows that the FoundationTTS adaptation obtains 40.13\% WER reduction when compared with the RNN-T baseline, and gets 10.35\% WER reduction compared to FastSpeech adaptation, which indicates that FoundationTTS can synthesize more diverse and realistic audio than traditional non-autoregressive continuous acoustic frames based TTS model, the evaluation set contains 1032 sentences. The adaptation experiment on hybrid and RNN-T ASR models shows that the synthesized data is robust enough for different ASR models which consistently achieve WER reduction.

\begin{table}[h]
  \centering
  %\resizebox{\linewidth}{!}{%
  \begin{tabular}{lll}
    \toprule
    Method & WER \\
    \midrule
    RNN-T Baseline & 15.05\% & \\
    \midrule
    +FastSpeech adaptation & 10.05\% & \\ 
    \midrule
    +FoundationTTS adaptation & 9.01\% & \\ 
    \bottomrule
  \end{tabular}
  \caption{WER Comparison of RNN-T, FastSpeech, and FoundationTTS on English adaptation test set. }
  \label{tab:RNN-T}
\end{table}

\begin{table}[h]
  \centering
  %\resizebox{\linewidth}{!}{%
  \begin{tabular}{lll}
    \toprule
    System & MOS \\
    \midrule
    FastSpeech & 3.84 ± 0.09 & \\
    \midrule
    FoundationTTS & 3.98 ± 0.08 & \\
    \midrule
    Recording & 4.45 ± 0.10 & \\
    \bottomrule
  \end{tabular}
  \caption{MOS Comparison of FoundationTTS vs FastSpeech trained on 900 hours diverse corpus.}
  \label{tab:mos}
\end{table}

\subsubsection{TTS Quality Evaluation}

\noindent \textbf{Subjective metrics } Table~\ref{tab:mos} shows the comparison between FoundationTTS and FastSpeech by the metrics of MOS. FoundationTTS and FastSpeech are trained with the same training data that consists of 900 hours of English corpus and more than 7000 speakers. The evaluation set contains 80 sentences, and each sample is rated by 10 judges, Results show that FoundationTTS outperforms FastSpeech in terms of naturalness. 

\noindent \textbf{Objective metrics of codec} We report the ViSQOL score of fine-grained codec and coarse-grained codec in Table~\ref{tab:obj1}. The evaluation set contains 2939 sentences. The bitrates of fine-grained codec and coarse-grained codec are 3.42 and 0.27 respectively. We can observe that ViSQOL score of fine-grained codec is +0.7 over coarse-grained codec, but coarse-grained codec using only 7.9\% bitrates compared with fine-grained codec, higher bitrates of fine-grained codec lead to better audio reconstruct quality and lower bitrates of coarse-grained codec results in shorter discrete sequences, which is more suitable for large language modeling. The combination of fine-grained codec and coarse-grained codec balances the reconstruction quality and modeling efficiency. Note that we have tried to train the coarse-grained codec directly on the waveform, but the model fails to converge. This implies that the fine-grained codec simplifies the feature representation of the audio waveform and makes the coarse-grained codec easier to train.

\noindent \textbf{Subjective metrics of codec} We also report CMOS score between audio reconstructed by fine-grained codec and recording, as shown in table \ref{tab:cmos}. The overall mean score of -0.06 indicates that the fine-grained codec can resynthesize high-fidelity speech waveform.

\begin{table}[h]
  \centering
  %\resizebox{\linewidth}{!}{%
  \begin{tabular}{lll}
    \toprule
     Codec & Bitrate (kbps) & ViSQOL \\
    \midrule
    Coarse-grained Codec & 0.27 & 3.07 \\
    \midrule
    Fine-grained Codec & 3.42 & 3.77 \\
    \bottomrule
  \end{tabular}
  \caption{Comparison of ViSQOL for hierarchical audio codec.}
  \label{tab:obj1}
\end{table}

\begin{table}[h]
  \centering
  %\resizebox{\linewidth}{!}{%
  \begin{tabular}{ll}
    \toprule
    System & CMOS \\
    \midrule
    Codec Reconstruction vs Recording & -0.06 \\
    \bottomrule
  \end{tabular}
  \caption{CMOS for fine-grained codec reconstruction.}
  \label{tab:cmos}
\end{table}

\subsubsection{Ablation Analysis}

\noindent Parameter number affects the stability and audio quality for the language model, we trained a series of speech token language models with different sizes ranging from 20M to 600M, which shows that the number of parameters significantly affects the end-to-end quality, the smallest 20M model cannot converge at all. Larger and more diverse training data requires a bigger model capacity, like what the GPT series model has shown in language.

The coarse-grained codec uses one quantizer in this work. We experiment with different quantizer numbers for the coarse-grained codec, which shows more quantizer has more audio reconstruction quality, but one speech frame will have more token id which is not suitable for language modeling, especially for the nosier dataset, less quantizer coarse codec is hard to better reconstruct the audio with high SNR or prosody varied corpus.

\section{Conclusions}

This paper describes FoundationTTS, a neural Text-to-Speech system that combines a hierarchical codec network with adversarial vector-quantized auto-encoders on continuous speech frames and a language model on discrete speech tokens, providing diverse speech output with large-scale and complex real speech dataset modeling ability without pre-designed acoustic features like mel or linear scale spectrograms. FoundationTTS achieves better quality than baseline systems on two ASR adaptation tasks. For future work, we plan to investigate a larger dataset with LM on more challenging datasets such as multi-speaker or multilingual datasets to explore the challenges and the potential application for realistic speech syntheses scenarios and ASR adaptation.

FoundationTTS is capable of synthesizing high-quality speech that closely matches the naturalness and speaker similarity of human speech, with diverse and smooth prosody. However, this advanced capability also comes with potential risks, such as voice cloning of specific individuals, which could be misused. To address these risks, we are committed to implementing Microsoft AI Principles in the further development of these models for various scenarios.\footnote{\url{https://www.microsoft.com/ai/responsible-ai}}.

\bibliographystyle{IEEEtran}

\bibliography{mybib}

\end{document}